\pdfoutput=1

\documentclass{article}

\usepackage[utf8]{inputenc} 
\usepackage[T1]{fontenc}    
\usepackage{hyperref}       
\usepackage{url}            
\usepackage{booktabs}       
\usepackage{amsfonts}       
\usepackage{nicefrac}       
\usepackage{microtype}      
\usepackage{color}
\usepackage[dvipsnames]{xcolor}
\usepackage{graphicx}
\usepackage{floatrow}
\usepackage{subfig}
\usepackage[percent]{overpic}
\usepackage{amsmath}
\usepackage{amssymb}
\usepackage{enumitem}
\usepackage{ulem}
\usepackage{comment}
\usepackage{cleveref}
\usepackage{footmisc}
\usepackage{notoccite}
\usepackage{soul}

\hypersetup{
  colorlinks,
  citecolor=Blue,
  linkcolor=Blue,
  urlcolor=Blue}

\usepackage[preprint]{neurips}

\DeclareFloatVCode{afterfloat}{}
\title{Towards falsifiable interpretability research}

\newif\ifshow
\showtrue
\showfalse

\ifshow
\newcommand{\ari}[1]{\textcolor{blue}{[\textbf{Ari:} #1]}}
\newcommand{\mat}[1]{\textcolor{Aquamarine}{[\textbf{Mat:} #1]}}
\newcommand{\todo}[1]{\textcolor{red}{[TO DO: #1]}}
\else
\newcommand{\ari}[1]{}
\newcommand{\mat}[1]{}
\newcommand{\todo}[1]{}
\fi

\newcommand*{\fullref}[1]{\hyperref[{#1}]{\autoref*{#1} \nameref*{#1}}}
\DeclareUnicodeCharacter{2212}{-}

\newcommand*\samethanks[1][\value{footnote}]{\footnotemark[#1]}

\author{%
  Matthew L. Leavitt\thanks{Equal contribution} \thanks{Work performed as part of the Facebook AI Residency program} \\
  Facebook AI Research\\
  Menlo Park, California\\
  \texttt{ito@fb.com} \\
   \And
   Ari S. Morcos\samethanks[1] \\
   Facebook AI Research \\
   Menlo Park, California \\
   \texttt{arimorcos@fb.com} \\
}

\begin{document}

\maketitle

\begin{abstract}

Methods for understanding the decisions of and mechanisms underlying deep neural networks (DNNs) typically rely on building intuition by emphasizing sensory or semantic features of individual examples. For instance, methods aim to visualize the components of an input which are “important” to a network’s decision, or to measure the semantic properties of single neurons. Here, we argue that interpretability research suffers from an over-reliance on intuition-based approaches that risk—and in some cases have caused—illusory progress and misleading conclusions. We identify a set of limitations that we argue impede meaningful progress in interpretability research, and examine two popular classes of interpretability methods—saliency and single-neuron-based approaches—that serve as case studies for how overreliance on intuition and lack of falsifiability can undermine interpretability research. To address these concerns, we propose a strategy to address these impediments in the form of a framework for strongly falsifiable interpretability research. We encourage researchers to use their intuitions as a starting point to develop and test clear, falsifiable hypotheses, and hope that our framework yields robust, evidence-based interpretability methods that generate meaningful advances in our understanding of DNNs.

\end{abstract}
\vspace{-4mm}
\section{Introduction} \label{sec:introduction}
\begin{quotation}
    \textit{In so far as a scientific statement speaks about reality, it must be falsifiable: and in so far as it is not falsifiable, it does not speak about reality.}
    
    -Karl Popper
    
    \textit{The Logic of Scientific Discovery} \citep{popper_logic_2005}
\end{quotation}

The complexity and high-dimensionality of deep neural networks (DNNs) makes them challenging to understand. One approach for understanding such systems is to extrapolate from intuitions about simpler, lower-dimensional systems. Mapping abstract quantities into concrete sensory forms, such as by visualization, can also be a powerful cognitive aid. However, looks are often deceiving and unchecked extrapolation can easily lead one astray.

The Gaussian (normal) distribution provides a vivid example of the pitfalls of extrapolating intuitions about low-dimensional systems to higher dimensions. A one-dimensional Gaussian distribution is perhaps the most familiar data distribution. Its iconic bell-like silhouette is ubiquitous across textbooks, technical publications, and lay reports of polls and statistics. Its one-dimensional shape extrapolates intuitively into two-dimensions: similar to an idealized mountain on a relief map. One might then expect that the “shape” of a Gaussian remains consistent as its dimensionality increases, with the majority of its mass lying close to the origin, with mass falling off like a bell curve as distance from the origin increases. Counterintuitively, however, most of the mass of a high-dimensional Gaussian is concentrated in a thin band at a constant distance from the origin. Thus the closest analogy for the shape of a high-dimensional Gaussian distribution is that of a soap bubble, or spherical shell: a fundamentally different structure than that of low-dimensional Gaussians \citep{blum_foundations_gaussian_shell_2020,huszar_gaussian_blog_2017, vershynin2018high}. 

Intuition is the foundation upon which comprehensive understanding is built. But, as the example of the Gaussian distribution shows, unverified intuition can be misleading. The basis of one’s intuition must be verified at every iteration of extrapolation. While intuition helps reveal what the important questions are, those questions need to be addressed with strong, falsifiable hypotheses. This applies to all science, including deep learning.

\mat{modified the final couple sentences of this paragraph}The nonlinearity of DNNs renders them resistant to being described and understood using analytical methods, leading researchers to rely instead on empirical approaches. A breadth of such approaches have been developed, typically referred to using the umbrella terms “interpretability” or “explainability” \citep{yosinski_understanding_2015, bau_network_2017, olah_building_blocks_2018, donnelly_interpretability_2019, radford_sentiment_neuron_2017, erhan_visualizing_2009, zeiler_deconvolution_2014, karpathy_visualizing_rnns_2016, amjad_understanding_2018, dhamdhere_conductance_2019, olah_zoom_2020, olah_feature_visualization_2017, zhou_object_2015, zhou_network_dissection_2019, gunning_explainable_2017, samek_explainable_2017, dosilovic_explainable_2018,lakkaraju_interpretable_2017,vellido_making_interpretability_2012,cadamuro_debugging_interpretability_2016,montavon_methods_interpretability_2018,guidotti_survey_interpretability_2018,lipton_mythos_interpretability_2018, hohman_visual_analytics_2019, roscher_explainable_2020}. Many of these methods rely on building intuition by emphasizing sensory or semantic features of individual examples, for instance by visualizing the components of an input which are “important” to a network’s decision \citep{springenberg_striving_saliency_2015,selvaraju_grad-cam_saliency_2017,fong_interpretable_saliency_2017,baehrens_how_saliency_2010,simonyan_deep_saliency_2014,zeiler_deconvolution_2014,shrikumar_learning_2017,sundararajan_axiomatic_saliency_2017,smilkov_smoothgrad_saliency_2017,dabkowski_real_time_saliency_2017,ancona_towards_saliency_2018}, or characterizing semantic properties of individual DNN neurons \citep{amjad_understanding_2018, meyes_ablation_2019, morcos_single_directions_2018, zhou_object_2015, zhou_revisiting_2018, bau_network_2017, karpathy_visualizing_rnns_2016, na_nlp_concepts_single_units_2019, radford_sentiment_neuron_2017, rafegas_indexing_selectivity_2019}. However, these practices make two assumptions that we argue are too frequently unverified in interpretability research: the assumption that a method faithfully represents meaningful properties of the model it is applied to, and the assumption that individual instances or examples faithfully represent the phenomena they are being used to illustrate. And their lack of verification is symptomatic of a more general lack of empirical rigour that impedes interpretability research. 

\mat{changed}In this position paper we discuss a set of impediments to robust interpretability research: a lack of concrete, falsifiable hypotheses; the abundance and seductive allure of visualization; under-verification of the principles upon which interpretability methods are based; a lack of quantifiable approaches; and a lack user research for human verification (Section \ref{sec:impediments}). We next review two case studies in the literature which exemplify the pitfalls of pursuing un(der)verified intuitions in deep learning, and perform a "post-mortem" to identify and explain the impediments that led to misleading conclusions (Sections \ref{sec:case1_saliency} and \ref{sec:case2_single_units}). We then articulate a strategy to address these impediments in the form of a framework for robust, falsifiable interpretability research (Section \ref{sec:best_practices}). And we finish by applying our framework to strengthen an example hypothesis (Section \ref{sec:example_hypotheses}).

\mat{new paragraph, who dis?}The motivation for this paper is not unprecedented; pleas have previously been made for greater rigor in the conduct and communication of machine learning research more broadly \citep{lipton_troubling_2018,forde_scientific_2019}, and in interpretability in particular \citep{doshi-velez_towards_2017,lipton_mythos_interpretability_2018}. Both \citet{lipton_mythos_interpretability_2018} and \citet{doshi-velez_towards_2017}, motivated by a lack of consensus and rigor regarding the definition, aims, and practices of interpretability research, sought to clarify these factors. In the intervening time since these publications, the technical shortcomings of the approaches in our case studies have precipitated, and many of them have been amended or alternative approaches have been developed. We think it's important not just to amend theses technical shortcomings, but also to analyze them so that we can generalize across cases to identify the impediments that allowed these technical shortcomings to arise in the first place, and determine how we can conduct interpretability research in a manner that avoids these impediments. As such, the recommendations in this paper focus more specifically on interpretability practices and less on the definitions and aims that were key parts of \citet{lipton_mythos_interpretability_2018} and \citet{doshi-velez_towards_2017}. Our core aim is to articulate concrete ways of "doing good science" in the context of interpretability research, by way of referencing both exemplary techniques and cautionary tales. \ari{I think this is good, but I would modify to emphasize that we also provide clear examples of the dangers of not being rigorous or related}

We emphasize that we are not arguing that intuition should be dispensed with altogether; intuition is essential for building understanding, and all else being equal, an intuitive method is strictly preferable to an unintuitive method. We are instead arguing that \textit{unverified} intuition and other aspects of interpretability research can facilitate misapprehension. Accordingly, the goal of this paper is to prevent misdirected effort, better realize the potential of useful approaches, and facilitate more impactful research through the application of scientific rigor.
\section{Impediments to falsifiable interpretability research} \label{sec:impediments}

Here, we discuss a set of impediments to interpretability research whose resolution would improve the robustness and reliability of our methods and the knowledge they generate. The primary impediment is an underemphasis on clear, specific, testable, and most critically, falsifiable hypotheses. This includes a lack of concrete null and alternative outcomes in the formulation of interpretability-guided hypotheses. We note that an interpretability method itself doesn’t need to be used to test a hypothesis directly; it can provide an intuition that forms the basis of a hypothesis that is testable with other methods.

Not only are interpretability methods underutilized for hypothesis formulation, but \textit{the methods themselves are too often left insufficiently scrutinized}. Specifically, they’re not verified as being important for DNN function—for example the assumption that some property of a single neuron is consequential to the behavior of the network. We believe that this issue is exacerbated by the innate visual appeal of many methods, which can provide a potentially false sense of comprehension, serving as an accelerant for unverified approaches.

\textit{The abundance and seductive allure of visualization} can be another impediment to robust interpretability research. Finding compelling, intuitive ways of representing information is critical for building understanding \citep{tufte1983visual,victor2014humane}, and many interpretability methods produce rich, stimulating visualizations \citep{mordvintsev_deepdream-code_2015,mordvintsev_inceptionism_2015,gatys_style_transfer_2016,olah_building_blocks_2018,olah_feature_visualization_2017,olah_zoom_2020,gatys_neural_2015,zeiler_deconvolution_2014}. But the illustrative power of visualization is a double-edged sword: an evocative graphic can elicit a strong feeling of comprehension regardless of whether the graphic faithfully represents the phenomenon it is attempting to depict. This makes interpretability a domain in which it is particularly easy to fall prey to unverified intuitions, and emphasizes the importance of faithful visualization.  

\textit{Lack of quantification} also impedes interpretability research. While not all methods can be quantified, it's hard to move beyond intuition without quantification. Quantification relates directly to the issue of verification, because quantification is often essential for verification. Anecdotal inspection of visualizations is not a method that scales to the number of scenarios that need to be evaluated to robustly verify a phenomenon. The number of models, data samples, hyperparameter configurations, and so on—the very factors that make deep learning effective—make this untenable. Furthermore, visual inspection presents risks. It can lead to cherry picking—intentionally or unintentionally—samples or instances that over-emphasize certain features or properties. It can also serve as a Rorschach test for a researcher, reinforcing their biases and adding unnecessary researcher degrees of freedom \citep{simmons_researcher_degrees_freedom_2011}.\footnote{Researcher degrees of freedom are described by \citet{simmons_researcher_degrees_freedom_2011} as follows: ``In the course of collecting and analyzing data, researchers have many decisions to make: Should more data be collected? Should some observations be excluded? Which conditions should be combined and which ones compared? Which control variables should be considered? Should specific measures be combined or transformed or both?'' All of these factors can encourage misleading conclusions.\label{fn:researcher_dof}}

The final impediment regards the importance of user research for human verification. Despite some interpretability methods being designed with the stated intent to provide actionable explanations to humans, many studies neglect to conduct \textit{controlled experiments using human subjects} to evaluate the quality and utility of these methods.


\section{Case study: The misdirection of saliency}
\label{sec:case1_saliency}

``Saliency'' methods (also known as feature attribution methods) refer to a collection of approaches that attempt to provide the user with insight as to “why” a DNN made a specific decision (i.e. performed an instance of inference) by identifying the features or parts of the input that are most relevant (or ``salient'') to the model's output, typically by using some version of the gradient of the output with regard to the input \citep{springenberg_striving_saliency_2015,selvaraju_grad-cam_saliency_2017,fong_interpretable_saliency_2017,baehrens_how_saliency_2010,simonyan_deep_saliency_2014,zeiler_deconvolution_2014,shrikumar_learning_2017,sundararajan_axiomatic_saliency_2017,smilkov_smoothgrad_saliency_2017,dabkowski_real_time_saliency_2017,ancona_towards_saliency_2018, adebayo_sanity_saliency_2018}. Saliency methods generate rich, visually-appealing annotations or transformations of input samples. For example, a saliency method applied to a network trained for object recognition might generate a heatmap across an example image of a cat that highlights the cat’s ears, leading to the conclusion that ears are the most salient aspect of the image for the network’s determination of ``catness''. Such a rich yet straightforward visualization strongly appeals to intuition. Indeed, the widespread study and application of saliency methods speaks to their intuitive appeal \citep{springenberg_striving_saliency_2015,selvaraju_grad-cam_saliency_2017,fong_interpretable_saliency_2017,baehrens_how_saliency_2010,simonyan_deep_saliency_2014,zeiler_deconvolution_2014,shrikumar_learning_2017,sundararajan_axiomatic_saliency_2017,smilkov_smoothgrad_saliency_2017,dabkowski_real_time_saliency_2017,ancona_towards_saliency_2018, yamashita_med_saliency_review_2018, rajpurkar_med_saliency_chexnet_2017, wang_med_saliency_chestx-ray8_2017, wang_saliency_rl_2020, puri_explain_saliency_rl_2019, mott_towards_rl_saliency_2019, wang_dueling_rl_saliency_2016, iyer_transparency_rl_saliency_2018, greydanus_visualizing_rl_saliency_2018, nikulin_free-lunch_rl_saliency_2019}.

But recent studies have cast doubt on their reliability \citep{alqaraawi_human_saliency_2020, kindermans_reliability_saliency_2017, sundararajan_axiomatic_saliency_2017, binder_layer-wise_2016, shrikumar_learning_2017, chandrasekaran_two_tango_2017, adebayo_sanity_saliency_2018, wang_saliency_skeptical_2019, atrey_exploratory_saliency_debunk_2019}. \citet{sundararajan_axiomatic_saliency_2017, binder_layer-wise_2016, shrikumar_learning_2017} identified that many studies of saliency methods lack a clear baseline for comparison (i.e. null hypothesis), a necessary condition for robust interpretability research. 
The notice of its absence presaged larger issues with saliency methods.

For saliency methods to be useful, they need to reflect meaningful properties of both the data and the networks to which they’re applied. For example, if an input and a baseline (i.e. control image) differ in one feature and the network has different predictions for the input and baseline, then the different feature should be identified as salient; likewise if a DNN does not depend on some feature, that feature should have zero salience. \citet{sundararajan_axiomatic_saliency_2017} identified that many saliency methods fail to exhibit this pair of traits, which they term “sensitivity”. The lack of sensitivity indicates that some saliency methods are prone to false negative and false positive assessments of the relevance of input features.

Saliency methods also need to reflect invariances of the networks to which they’re applied. Take the example of networks with different architectures or parameters but identical output behavior across a dataset, which we term to be functionally equivalent. If two networks are functionally equivalent in this manner, a saliency method should generate identical results for the two networks. \citet{sundararajan_axiomatic_saliency_2017} refer to this as implementation invariance, and showed that many saliency methods do not exhibit this property. This concept can also be extended to the input data: saliency methods should be invariant to input transformations that do not affect model predictions or weights, a property that \citep{kindermans_reliability_saliency_2017} found is not obeyed by many saliency methods. These results indicate that many saliency methods do not reflect meaningful properties of the systems they are used to understand.

Perhaps most importantly, saliency methods are intended to help understand DNNs. Accordingly, they are only useful for this purpose if they provide information about the data with regards to the model, not just the data in isolation. \citet{adebayo_sanity_saliency_2018} examined a suite of saliency methods and found that many of them produced nearly identical saliency maps \textit{even when all the model’s parameters were reinitialized or the models were trained on label-shuffled data}. They also used an edge detector as an additional control, because edge detectors are not specific to either models or training data, and found that the outputs of the saliency methods were in many cases very similar to those of the edge detector. Because gradient-based methods naturally emphasize edges, and because edges are inherently salient to humans \citep{henderson_human_2003,geisler_visual_2008,bansal_edges_2013}, this means that saliency methods will likely generate maps which appear consistent with human experience, regardless of whether they reflect meaningful properties of the model, undermining their primary purpose of providing insight about a model’s decision. Put another way, these methods provided the answer we wanted to hear regardless of whether or not it was true. 


Saliency methods are intended to provide humans with an understanding of ``why'' a DNN made a decision. Accordingly, providing humans with the results of saliency methods should improve their ability to reason about and/or predict the behavior of DNNs. \citet{chandrasekaran_two_tango_2017} used a Visual Question Answering (VQA) task and found that providing humans with saliency maps for the image and/or words in a data sample do not improve subjects’ predictions of the DNN’s behavior. A subsequent study \citep{alqaraawi_human_saliency_2020} found that saliency maps only weakly improve participants’ abilities to anticipate a network’s output for new images. These studies call into question the efficacy of saliency for achieving its stated goal and highlight the necessity of verifying the human side of interpretability methods. It is possible that saliency methods lack utility for aiding human predictions because they generate outputs that appear meaningful to humans but are unrelated to the models, indicating that rigorous methodological controls and human verification are two sides of the same coin. 

It is unfortunate that many saliency methods have not held up to scrutiny, especially given their application to extremely sensitive settings such as medical imaging \citep{seo_med_saliency_visualizing_2019,amit_med_saliency_hybrid_2017,wang_med_saliency_chestx-ray8_2017,wang_med_saliency_tienet_2018,zhang_med_saliency_mdnet_2017,mahapatra_med_saliency_retinal_2016,rajpurkar_med_saliency_chexnet_2017,yamashita_med_saliency_review_2018,karargyris_med_saliency_age_2019}, which present a significant safety risk if deployed uncritically. Thankfully, there has been meaningful progress in assessing and developing robust saliency methods. \citet{kindermans_reliability_saliency_2017,adebayo_sanity_saliency_2018, arras_evaluating_2019, yeh_fidelity_2019, hooker_benchmark_2019, sanchezlengeling2020attribution} in particular provide guidance for verifying interpretability methods\footnote{It bears mentioning that some studies identifying limitations in saliency methods used their discoveries to design newer, more robust saliency methods, which were then in turn shown to have significant limitations. In our view, this demonstrates the critical roles of iteration and replication in scientific research.} and/or develop robust methods, and the work of \citet{chandrasekaran_two_tango_2017,doshi-velez_towards_2017,alqaraawi_human_saliency_2020} provide examples of how to evaluate human-relevance.

\section{Case study: Single units and (sub)optimal examples}
\label{sec:case2_single_units}

While saliency seeks to explain how a model made a decision with regard to a single input, our next case study examines a group of interpretability methods that aim to build intuitions for the broader mechanisms by which DNNs function \citep{zhou_object_2015, rafegas_indexing_selectivity_2019, erhan_visualizing_2009, zeiler_deconvolution_2014, simonyan_deep_saliency_2014, yosinski_understanding_2015, nguyen_preferred_synthesis_2016, olah_feature_visualization_2017, olah_building_blocks_2018, amjad_understanding_2018, meyes_ablation_2019, morcos_single_directions_2018, zhou_object_2015, zhou_revisiting_2018, bau_network_2017, karpathy_visualizing_rnns_2016, na_nlp_concepts_single_units_2019, radford_sentiment_neuron_2017, rafegas_indexing_selectivity_2019, olah_zoom_2020, nguyen_multifaceted_visual_2016, dosovitskiy_inverting_visual_2016, mahendran_understanding_visual_2014}. DNNs are extremely high-dimensional systems, and as the opening example with the Gaussian distribution demonstrated, it is challenging to mentally navigate high-dimensional spaces. This challenge is a strong motivator for humans to abandon high-dimensional thinking in favor of the familiarity and comfort of low-dimensional signals. This can in turn encourage the properties of individual units to be deployed as an explanation of ``what'' a network is ``doing'', which assumes the unit is both representative of other units and responsible for some delineated sub-component of the network’s behavior.  Many of these methods make claims based solely on visualization or qualitative properties \citep{zhou_object_2015, rafegas_indexing_selectivity_2019, erhan_visualizing_2009, zeiler_deconvolution_2014, simonyan_deep_saliency_2014, yosinski_understanding_2015, nguyen_preferred_synthesis_2016, olah_feature_visualization_2017, olah_building_blocks_2018, olah_zoom_2020, nguyen_multifaceted_visual_2016, dosovitskiy_inverting_visual_2016, mahendran_understanding_visual_2014, hohman_visual_analytics_2019}. We start by examining how our falsifiability framework applies to visualizations.

Visualization is a common method for building intuition about the properties of single units. In the most basic approach, one simply identifies the input sample(s) (e.g. images) or sample subregions that maximally activate a given neuron \citep{zhou_object_2015, rafegas_indexing_selectivity_2019}. Another approach uses optimization to generate samples that maximize unit activations \citep{erhan_visualizing_2009, zeiler_deconvolution_2014, simonyan_deep_saliency_2014, yosinski_understanding_2015, nguyen_preferred_synthesis_2016, olah_feature_visualization_2017, olah_building_blocks_2018, olah_zoom_2020, nguyen_multifaceted_visual_2016, dosovitskiy_inverting_visual_2016, mahendran_understanding_visual_2014}. While these methods may be useful for building intuition, they can also encourage three potentially misleading assumptions: that the visualization is representative of the neuron’s behavior; that the neuron is responsible for a clearly delineated portion of the task or the network’s behavior; and that the neuron’s behavior is representative of the network’s behavior.

To illustrate how visualization can provide misleading intuitions about a neuron’s properties, we summarize \citet{huszar_gaussian_blog_2017}'s lucid example: If a neuron responds to images of Gaussian noise, a method that generates the “optimal” input for that neuron may end up generating a gray square, which is very different from the samples it most strongly responds to: white noise images which look like static. \citep{olah_feature_visualization_2017,nguyen_multifaceted_visual_2016} are also motivated by the potential of single instances to generate misleading conclusions. 

In addition to visualization methods potentially being misleading, they are often under-utilized for informing and evaluating strong, falsifiable hypotheses. While demonstrating that an optimal sample exists or that an optimal sample exhibits certain qualities are technically falsifiable hypotheses, these are not strong hypotheses. Stronger hypotheses would relate to network function instead of just proof of existence. It seems that visualization methods have a dual deficit when it comes to testing: their “representativeness” is not always tested, nor are they used to guide or test hypotheses.

Non-visualization-based methods for characterizing the behavior of individual neurons—typically by characterizing the distribution of the neuron’s activity over data classes or values of a data feature—are myriad and ubiquitous \citep{amjad_understanding_2018, meyes_ablation_2019, morcos_single_directions_2018, zhou_object_2015, zhou_revisiting_2018, bau_network_2017, karpathy_visualizing_rnns_2016, na_nlp_concepts_single_units_2019, radford_sentiment_neuron_2017, rafegas_indexing_selectivity_2019}. Because the practice is imported from neuroscience \citep{sherrington_integrative_1906, adrian_impulses_1926, granit_receptors_1955, hubel_receptive_1959, barlow_single_1972, kandel_principles_2000}, we will abide by their lexicon and collectively refer to these practices as measuring “selectivity”. Selectivity-based methods have a similar allure as visualization, but the appeal is semantic instead of sensory. One notable study identified a neuron selective for sentiment in an LSTM network trained on a word prediction task \citep{radford_sentiment_neuron_2017}. Another found individual neurons selective for cats faces and human bodies \citep{radford_sentiment_neuron_2017}. \citep{olah_feature_visualization_2017,olah_building_blocks_2018,olah_zoom_2020} propose single unit selectivity as a building block for understanding DNNs. The influence of these studies demonstrates the widespread appeal of selectivity \citep{amjad_understanding_2018, meyes_ablation_2019, morcos_single_directions_2018, zhou_object_2015, zhou_revisiting_2018, bau_network_2017, karpathy_visualizing_rnns_2016, na_nlp_concepts_single_units_2019, radford_sentiment_neuron_2017, rafegas_indexing_selectivity_2019, olah_feature_visualization_2017, olah_building_blocks_2018, olah_zoom_2020}. But neither semantic abstraction nor quantification completely avoid the pitfalls of visualization, because they can still encourage the assumptions that a neuron is responsible for a clearly delineated portion of the task or the network’s function, or is representative of a network’s behavior.

A number of recent studies examining class selectivity—a form of selectivity defined as how variable a neuron’s responses are across different classes of data samples—have demonstrated that the properties of single neurons may not robustly account for the behavior of networks \citet{morcos_single_directions_2018} used single unit ablation and other techniques to demonstrate that a network's test set generalization is negatively correlated (or uncorrelated) with the class selectivity of its units, suggesting that highly-selective units are no more important (or perhaps even less important) than confusing, non-selective units. \citet{amjad_understanding_2018} confirmed these results for single unit ablation, but also performed cumulative ablation analyses where multiple units are ablated, the results of which suggested that selectivity is mildly beneficial. This dichotomy presents one example of how it can be difficult to extrapolate from single neurons to networks, in this case because of redundancy across units. 

Similar results have been shown for models trained on NLP tasks. \citet{dalvi_grain_of_sand_2019} found that ablating units selective for linguistic features causes greater performance deficits than ablating less-selective units. \citet{donnelly_interpretability_2019} found that ablating \citet{radford_sentiment_neuron_2017}’s ``sentiment neuron'' has minimal impact on the network’s ability to detect and label sentiment. These results certainly cast doubt on the assumptions that individual neurons contribute to clearly delineated portions of the network’s behavior and that individual neurons are representative of the network, but they are also limited by their reliance on the method of single unit ablation.

\mat{edited}As detailed in \citet{leavitt_selectivity_2020}, single unit ablation in trained networks can only address the effects of inactivating a neuron in a network whose training process assumed that the neuron was present. This presents two key limitations to investigating the causal role of selectivity: single unit ablation is unable to address whether the presence of selectivity is beneficial (i.e. is it sufficient), and it is unable to address whether networks need to learn selectivity to function properly. With these limitations in mind, \citet{leavitt_selectivity_2020} conducted a study in which they regularized against class selectivity in the loss function—sidestepping the shortcomings of single unit ablation—to examine the causal role of class selectivity in network function. They found that class selectivity is largely unnecessary for network function, and reducing it can even improve network performance. They also found in a follow-up study that reducing class selectivity increases robustness to naturalistic corruptions of input data \citep{leavitt2020relationship}. These results demonstrate that networks do not need to learn class selectivity, and that in some cases learning class selectivity can impair network function, indicating that examining class selectivity in single neurons may be a red herring with regards to understanding network function.

\mat{modified}The studies we detailed support the notion that the tractable, semantic allure of single neurons’ low-dimensional representations can encourage an overreliance on intuition. And while these studies question the functional relevance of class selectivity, we view them more broadly as a bulwark against the intuitive appeal of focusing on single neuron properties. This suggests that researchers should focus on properties that exist across neurons—distributed, high-dimensional representations—and highlights the need to develop tools to make these properties more tractable and accessible to intuition. \citet{agrawal_analyzing_2014,wang_unsupervised_2016,alain_understanding_2018,morcos_cca_2018,fong_net2vec_2018} are examples of studies that attempt to better understand DNNs by examining distributed representations.\mat{wondering whether I should include any Olah's work here; many of the papers examined distributed representations in addition to single units} \mat{feel like this is a sour ending, but also feel like it's a good link to the next section} \ari{I think the point at the end of some of these sections is to highlight the good, so it feels weird to say, "these are good, but not that good". Let's just focus on the good ones I'd say if we can.} \ari{I think we can add a cite to the circuits work, though that's not super falsifiable, but maybe that's ok because that's not really the point of this section? This one is more about not quantifying. That said, obviously closely related.}

\section{Best practices for robust, falsifiable interpretability research} \label{sec:best_practices}

We have attempted to illustrate how the seductive appeal of intuition and a lack of verification and falsifiability have resulted in misleading conclusions and illusory progress in two domains of interpretability research: saliency and single neuron-based approaches. What follows is a high-level framework for best practices to help ensure robust, falsifiable interpretability research. These practices are guided by the impediments to interpretability research presented in Section \ref{sec:impediments}, and by the techniques and interventions that revealed the limitations of saliency and single unit-based approaches. We provide practical examples and references whenever possible. We would also like to note that it is not our intent to present this framework as a novel discovery, but simply as a clearly-articulated, practically-applicable collection of best practices.

Our first, and most important recommendation, is to make and test clear, specific, testable, and falsifiable hypotheses. This not only demonstrates or improves the impact of a method, but also serves as a sanity check for the efficacy of that method. However, this requires being able to dissociate whether a result is due to a property of the method or a property of the system being examined. For example, determining whether a saliency method is dependent on the properties of a model, or whether it functions more like an edge detector. It is also important to note that merely proving the existence of something rarely tells us much about whether that phenomenon is relevant to network function. Stronger hypotheses relate to causal mechanisms of network function, e.g., “perturbing the representation in dimension X should change the output in dimension Y”. \citet{bau_gan_dissection_2019} is a good example of this: their method, termed GAN dissection, identifies sets of units in GANs whose manipulation controls the presence of specific image features in the generated images, such as windows and people. A weaker alternative would have been, for example, to show that there exist certain subsets of units whose activity is correlated with the presence of certain features in the generated images. Furthermore, some hypotheses may be untestable using existing approaches. For example, single neuron ablation cannot address the sufficiency of class selectivity for network performance, nor whether class-selective solutions are necessary. The inability to address these causal hypotheses is what motivated \citet{leavitt_selectivity_2020} to develop their class selectivity regularizer.\mat{I added this as another example, and also to address Sara's point about mixed/inconclusive single neuron results}


Next, we caution researchers to be vigilant of the seductive allure of visualization. One’s skepticism should be proportional to the feeling of intuitiveness. Put simply: one should question their “wow”. 
If individual examples are being used to build intuition, they should be properly contextualized relative to other samples. For example, by providing a meaningful baseline for comparison \citep[e.g. edge detectors for saliency methods][]{adebayo_sanity_saliency_2018,hooker_benchmark_2019}, or by examining where the examples lie in the distribution(s) over the quantities of interest. This latter recommendation requires a correspondence between the visualizations and a quantifiable feature of the data or model, which leads to our next recommendation. 

\mat{updated with methods cites}We encourage researchers to quantify their methods. Not only does inspecting visualizations not scale to the size of deep learning models and datasets, but it can be dangerously prone to experimenter bias and adds researcher degrees of freedom\footref{fn:researcher_dof}. \citet{adebayo_sanity_saliency_2018}'s combination of two simple methods—randomization and the Spearman correlation—were sufficient to generate the bulk of their conclusions about the validity of different saliency methods.\footnote{They also used more sophisticated metrics, but primarily to buttress the findings obtained with the Spearman correlation.} And the challenge of quantifying saliency methods motivated \citet{kim_tcav_2018} to develop their Concept Activation Vector approach for interpretability; likewise, \citet{carlini_prototypical_2019,jiang_characterizing_2020,agarwal_estimating_2020} all propose and/or evaluate methods for quantifying the difficulty or prototypicality of data samples. While not all methods can be quantified, it is challenging to move beyond intuition without quantification: an unquantifiable hypothesis risks being an unfalsifiable hypothesis.

Our final recommendation is to remember the “human” in “human explainability”. Keep in mind whether a method is intended to inform a human, and verify that it does so. Deep learning researchers may not feel comfortable conducting human behavioral research, but \citet{chandrasekaran_two_tango_2017,doshi-velez_towards_2017,alqaraawi_human_saliency_2020,berardino_eigen-distortions_2017} illustrate straightforward approaches that yielded clear insights.
\section{Building better hypotheses}
\label{sec:example_hypotheses}
In this section, we demonstrate how hypotheses can be strengthened with respect to interpretability. We start with a very weak example hypothesis, identify its weaknesses, and then use this information to improve it. We iteratively apply this process until we obtain a strong hypothesis, identifying the strengths and weaknesses of each hypothesis along the way. For the sake of relevance our example hypotheses address the role of feature-selective neurons in DNN function, but our aim is to provide a generalizable example for how to build a strong scientific hypothesis.

\textbf{Very weak hypothesis:} \textit{Feature-selective neurons are the foundation of DNN function.}


The most significant weakness of this hypothesis is that it is not falsifiable. ``Foundation'' is vague, causing the hypothesis to lack a concrete alternative and testable predictions. Consequently, the best attempt to articulate an alternative hypothesis yields something as vague as the original: ``feature-selective neurons are not the foundation of DNN function''. What would it mean for feature-selective neurons not to be the ``foundation'' of DNN function? Determining the validity of this hypothesis seems to be more dependent on semantics than science.



\textbf{Weak hypothesis:} \textit{If feature selectivity is important for DNN function, then we should find feature-selective neurons.}

This hypothesis improves on the prior example by making a specific, falsifiable prediction: ``we should find feature-selective neurons''. Unfortunately, proving the existence of some phenomenon is typically a weak outcome that lacks explanatory power in the absence of sufficient context and/or strong pre-existing evidence against its existence. This hypothesis also provides no basis for evaluating the strength or validity of its prediction: Is the baseline just the number of feature-selective neurons expected to exist by chance? And is it meaningful to simply find a quantity of feature-selective neurons that exceeds the baseline? Is a baseline used all? Finally, this hypothesis lacks alternative potential explanations for the presence of feature-selective neurons: are there scenarios in which feature-selective neurons could exist even if feature selectivity is \textit{not} important for DNN function?





\textbf{Average hypothesis:} \textit{If feature selectivity is necessary to maximize test accuracy, ablating feature-selective single neurons should cause a decrease in test accuracy.}


This hypothesis improves on the prior examples by addressing causality: The term ``important'' has been replaced with ``necessary'', which specifies the causal role of feature-selective neurons. It also describes a specific experiment that is capable of addressing causality (ablation) and the predicted outcome of that experiment, which is clearly falsifiable and made in concrete terms (``DNN function'' becomes ``decrease in test accuracy''). 





\textbf{Strong hypothesis:} \textit{If feature selectivity in single neurons is necessary to maximize test accuracy, ablating neurons selective for a specific feature (e.g. curves) should cause a decrease in test accuracy proportional to the strength of the neuron’s feature selectivity for samples that contain that feature (e.g. round objects). Alternatively, if networks rely more on feature selectivity across neurons than on feature selectivity in individual neurons, then zero-ing activity in feature-selective directions (i.e. a linear combination of units that represents curves) that are not axis-aligned should cause a decrease in test accuracy that is proportional to the strength of feature selectivity and exceeds the decrease from ablating only single units.}

This formulation is substantially more comprehensive than any of the prior hypotheses. It dissociates necessity and sufficiency, and makes specific falsifiable predictions for these two factors. It also presents multiple competing hypotheses: the importance of single unit vs. distributed representations. The two experiments—ablating feature-selective single neurons and ablating feature-selective non-axis-aligned activity—allow the dissociation of necessity and sufficiency of these two forms of representations. A baseline for comparing the importance of the two forms of representation is also provided: the test accuracy decrease from ablating off-axis activity needs to exceed the test accuracy decrease from ablating single units for networks to be considered ``more reliant'' on distributed representations. 

\section{Conclusion}
\label{sec:conclusion}
We detailed four impediments to robust interpretability research arising from an overemphasis on building intuition: the abundance and seductive allure of visualization; under-verification of the principles upon which interpretability methods are based; a lack of quantifiable approaches; and under-utilization of user research for human verification. We then examined two classes of interpretability methods—saliency and single neuron-based approaches—in which these impediments yielded misleading results and undermined meaningful progress. We then synthesized these impediments into a framework for robust, falsifiable interpretability research, and finished by applying this framework to iteratively improve an example hypotheses. 

We would like to reiterate that the goal of this paper is to prevent misdirected effort, better realize the potential of useful approaches, and facilitate more impactful research through the application of scientific rigor. Intuition is an indispensable first step, but it has just as much power to mislead as it does to enlighten. By using these intuitions to develop and test clear, falsifiable hypotheses, we hope that researchers will converge on interpretability methods which are both intuitive \textit{and} stand up to scrutiny to reveal 
deep insights about DNNs. 
\subsubsection*{Acknowledgements} \label{sec:acknowledgements}
We would like to thank Been Kim and Sara Hooker for their insightful feedback.

\bibliography{ms}
\bibliographystyle{plainnat}



\end{document}